\documentclass[%
 reprint,
amsmath,amssymb,
aps,
]{revtex4-1}

\usepackage{graphicx}
\usepackage{dcolumn}
\usepackage{bm}

\begin{document}


\title{Sp($2$,$\mathbb{Z}$) invariant Wigner function on even dimensional vector space}

\author{Minoru Horibe}
\email{horibe@quantum.apphy.u-fukui.ac.jp}
\author{Takaaki Hashimoto}
\email{hasimoto@u-fukui.ac.jp}
\author{Akihisa Hayashi}
\email{hayashi@soliton.apphy.u-fukui.ac.jp}
\affiliation{%
Graduate School of Engineering, University of Fukui1 3-9-1, Fukui city, Fukui prefecture.
}%

\date{\today}

\begin{abstract}
We construct the quasi probability distribution $W(p,q)$ on even dimensional vector space with marginality and  invariance under
the transformation induced by projective representation of the group ${\rm Sp}(2,\mathbb{Z})$ whose elements  correspond to 
linear canonical transformation.

On even dimensional vector space, non-existence of such a 
quasi probability distribution whose arguments take physical values was shown in our previous paper(Phys.Rev.A{\bf 65} 032105(2002)).
For this reason
we study a quasi probability distribution $W(p,q)$ whose arguments $q$ and $p$ take not only $N$ physical values but also $N$ unphysical values, where $N$ is dimension of vector space. It is shown that there are two quasi probability distributions on even dimensional vector space. The one is equivalent to the Wigner function proposed by Leonhardt, and the other is
a new one.
\end{abstract}

\maketitle


\section{\label{sec:Intro}Introduction}

In quantum mechanics, as is well-known, we cannot define probability distribution function on phase space with which observed values for any operators are equal to the weighted average. At the expense of taking a negative value, we can define the function $W(q,p)$ which resembles probability distribution on
phase space. Such a function was presented by Wigner in 1932 \cite{wigner1} and called Wigner function, which is defined as:
\begin{equation}
W(q,p)={\rm Tr}\left[\hat{\Delta}(q,p)\hat{\rho}\right],
\label{wigfano}
\end{equation}
where $\hat{\rho}$ is a density matrix and $\hat{\Delta}(q,p)$ is the Fano operator given by
\[
\hat{\Delta}(q,p)=
\frac{1}{2\pi \hbar}\int_{-\infty}^{\infty}dr
\left|q-\frac{r}{2}\right\rangle
e^{-ipr/\hbar}
\left\langle q+\frac{r}{2} \right|.   
\]

It is easily checked that the Fano operator satisfies two properties, marginality
\begin{equation}
\begin{array}{l}
\displaystyle{\int_{-\infty}^{\infty}\hat{\Delta}(q,p)dp=|q\rangle \langle q|}, \\
\displaystyle{\int_{-\infty}^{\infty}\hat{\Delta}(q,p)dq=|p\rangle \langle p|},
\end{array}
\label{conmar1}
\end{equation}
and covariance under the linear canonical transformation,
\begin{eqnarray}
\hat{\Delta}'(q',p')&=&\hat{U}_{{\bf h}_{\rm c}}\hat{\Delta}(q',p')\hat{U}^{\dagger}_{{\bf h}_{\rm c}}\nonumber \\
&=&\hat{\Delta}(\delta q' - \beta p', -\gamma q' + \alpha p'), 
\label{contcov}
\end{eqnarray}
where $\hat{U}_{{\bf h}_{\rm c}}$ is unitary operator for linear canonical
transformation,
\begin{equation}
\left\{\begin{array}{l}
\hat{q}'=\hat{U}_{{\bf h}_{\rm c}}\hat{q}\hat{U}^{\dagger}_{{\bf h}_{\rm c}}
=\alpha \hat{q} + \beta \hat{p}, \\
\hat{p}'=\hat{U}_{{\bf h}_{\rm c}}\hat{p}\hat{U}^{\dagger}_{{\bf h}_{\rm c}}
=\gamma \hat{q} + \delta \hat{p},
\end{array}\right.  
\rightleftharpoons
\left\{\begin{array}{l}
\hat{q}= \delta \hat{q}' - \beta \hat{p}', \\
\hat{p}=-\gamma \hat{q}' + \alpha \hat{p},
\end{array}\right. \label{contran}
\end{equation}
and $2\times 2$ matrix ${\bf h}_{\rm c}$ is an element of group 
${\rm Sp}(2,\mathbb{R})$;
\begin{eqnarray*}
&&{\rm Sp}(2,\mathbb{R})\\
&&=\left\{{\bf h}_{\rm c},\biggl| 
 {\bf h}_{\rm c}=\left(\begin{array}{cc}
  \alpha & \gamma \\
  \beta  & \delta
  \end{array}\right),\alpha\delta-\beta \gamma=1,\;\;
   \alpha,\beta,\gamma,\delta \in \mathbb{R} \right\}.
\end{eqnarray*}
This covariance of the Fano operator means  the invariance of the Wigner function under changing the label of a point on phase space from $(q,p)$ to $(q',p')=(\alpha q + \beta p,\gamma q + \delta p)$ ;
\begin{eqnarray*}
W'(q',p')&=&{\rm Tr}\left[\hat{\Delta}'(q',p')\hat{\rho}\right]\\
         &=&{\rm Tr}\left[\hat{\Delta}(\delta q' - \beta p', -\gamma q' + \alpha p')
             \hat{\rho}\right]\\
         &=&W(\delta q' - \beta p',-\gamma q' + \alpha p')=W(q,p),
\end{eqnarray*}
so, we are mainly concerned with covariance of the Fano operator, rather than invariance for the Wigner function.

From marginality, we obtain the equation
\begin{equation}
\int_{-\infty}^{\infty}q^n\hat{\Delta}(q,p)dqdp=\hat{q}^n.
\label{conmar2}
\end{equation}
Applying the similarity transformation by the unitary operator $\hat{U}_{{\bf h}_{\rm c}}$ to both sides of eq.(\ref{conmar2}), we can find that the integration of the Fano operator multiplied by $q^np^m$ over phase space is equal to the Weyl ordering products 
$(q^np^m)_{\lower0.3ex\hbox{\tiny{\rm W}}}$ of
operator $\hat{q}^n\hat{p}^m$ which is given by
\begin{eqnarray*}
& &(\alpha \hat{q}+\beta \hat{p})^{K} \\
&=&\sum_{k=0}^{K}\left(\begin{array}{c}
K \\
 k
\end{array}\right)\alpha^{K-k}\beta^k(\hat{q}^{K-k}\hat{p}^k)_{\lower0.3ex\hbox{\tiny{\rm W}}}.
\end{eqnarray*}
This equation teaches us that the expectation value of Weyl ordered operator $(\hat{q}^n\hat{p}^m)_{\lower0.3ex\hbox{\tiny{\rm W}}}$ for
the density matrix $\hat{\rho}$ is equivalent to the 
``weighted average" of $q^np^m$ with respect to the Wigner function,
\[
{\rm Tr}[(\hat{q}^n\hat{p}^m)_{\lower0.3ex\hbox{\tiny{\rm W}}}\hat{\rho}]=\int_{-\infty}^{\infty}q^np^mW(q,p)dqdp.
\]
Conversely, in our previous paper\cite{horibe1}, it is shown that the Fano operator is uniquely determined if it is a continuous operator of $q$ and $p$ and satisfies the above two properties, namely,
marginality eq.(\ref{conmar1}) and covariance eq.(\ref{contcov}) under the linear canonical transformation. 
Other methods how to determine the Wigner function or the Fano operator uniquely are studied in literatures \cite{Li} and \cite{horibe2}. 

For the quantum system on the finite vector space, we investigated the same problems as on one-dimensional quantum system in the paper \cite{horibe1} and found that
the Wigner function with properties corresponding to the above two properties exists uniquely on odd dimensional vector space and that it does not exist on even dimensional vector space.

For even dimensional vector space, Leonhardt proposed the Wigner function $W(q,p)$ which resembles the one for the one-dimensional quantum system, although variables $q$ and $p$ take integers related to eigenvalues of corresponding operators and half-integers between $0$ and $N-\frac{1}{2}$ \cite{Leonhardt2}. 
We can check that the Fano operator $\hat{\Delta}^{({\rm L})}$ for this Wigner function satisfies
the covariance and marginality,
\begin{eqnarray}
\sum_{p\in {\cal IH}_N} \hat{\Delta}^{({\rm L})}(q,p)&=&
    \left\{
      \begin{array}{ll}
         |q \rangle \langle q| & (q \in {\cal I}_N) \\
             0                 & (q \in {\cal H}_N)
      \end{array}\right.,
\label{projcoord1}\\
\sum_{q\in {\cal IH}_N} \hat{\Delta}^{({\rm L})}(q,p)&=&
    \left\{
      \begin{array}{ll}
         |p \rangle \langle p| & (p \in {\cal I}_N) \\
             0                 & (p \in {\cal H}_N)
      \end{array}\right., \label{projmonent1}
\end{eqnarray}
where ${\cal I}_N$ and ${\cal H}_N$ are sets of integers and half-integers in right-open interval $[0,N)$, respectively,
\[
{\cal I}_N=\{0,1,2,\cdots, N-1\},\;\;
{\cal H}_N=\left\{\frac{1}{2},\frac{3}{2}, \cdots, \frac{2N-1}{2} \right\}.
\]
and the set ${\cal IH}_N$ is union of ${\cal I}_N$ and ${\cal H}_N$.

Can we determine the Fano operator on even dimensional vector
space if the above two conditions are imposed?
We study this problem in this note.

\section{\label{sec:preparation}Preparation}
As we cannot have a pair of operators which satisfy the usual
canonical commutation relation  for linear operators acting on finite dimensional vector space, we define operators which satisfy the
similar multiplication rule to one of
$e^{ia\hat{q}}$ and $e^{ib\hat{p}}$ ;
\[
e^{ia\hat{q}}e^{ib\hat{p}}=e^{-i\hbar ab}e^{ib\hat{p}}e^{ia\hat{q}}.
\] 
Using an orthogonal basis $\{|0 \rangle, \ |1 \rangle, \cdots, |N-1 \rangle\}$ of the vector space under consideration, 
the operators $\hat{Q}$ and $\hat{P}$ are defined by 
\begin{equation}
\begin{array}{l}
\displaystyle{\hat{Q}=\sum_{q \in {\cal I}_N}\omega_N^{q}|q\rangle\langle q|},\\
\displaystyle{\hat{P}=\sum_{p \in {\cal I}_N}\omega_N^{p}|p\rangle \langle p|},
\end{array}\label{defQP}
\end{equation}
where  $|p \rangle$ is the Fourier transformed vector of $|q\rangle$
\[
|p\rangle = \frac{1}{\sqrt{N}}\sum_{q\in {\cal I}_N}\omega_N^{pq}|q\rangle,
\]
and $\omega_N$ is a primitive $N$-th root of unity,
\[
\omega_N=e^{i\frac{2\pi}{N}}.
\]

In order to make equations simple, 
we assume that elements $|q \rangle $ and $ |p \rangle$   
satisfy periodical boundary conditions
\[
|q+N\rangle =|q \rangle,\;\;\;|p+N\rangle =|p \rangle.
\]
We can check that these operators to the $N$-th power are equal to unit matrix
\begin{equation}
\hat{Q}^N=\hat{P}^N={\bf 1}_N, \label{PQN} 
\end{equation}
and that $\hat{P}\hat{Q}$ is equal to $\hat{Q}\hat{P}$ multiplied by $\omega_N$
\begin{equation}
\hat{P}\hat{Q}=\omega_N\hat{Q}\hat{P} . 
\label{comQP}
\end{equation}

The operators $\hat{P}^m\hat{Q}^n,\;(n,m=0,1,2,\cdots,N-1)$ are linearly independent,
because of 
\[
{\rm Tr}[\hat{P}^k\hat{Q}^l(\hat{P}^m\hat{Q}^n)^{\dagger}]=
N\delta_{km}\delta_{ln}.
\]
Since the operator $\hat{E}_{qq'}=|q \rangle \langle q'|$ can be expressed with the expansion
\begin{equation}
|q \rangle \langle q'|=\frac{1}{N}\left\{\sum_{n \in {\cal I}_N}(\omega_N^{-q} \hat{Q})^{n}\right\}\hat{P}^{q'-q},
\label{expfunmat}
\end{equation}
any operators are decomposed into the series of $\hat{P}^n\hat{Q}^m,\;(n,m=0,1,2,\cdots,N-1)$
,uniquely. 
For example, the Fano operator $\hat{\Delta}^{({\rm L})}(q,p)$ for the Wigner function proposed by Leonhardt on even dimensional vector space
is expanded in the form,
\begin{eqnarray}
\hat{\Delta}^{({\rm L})}(q,p)&=&\frac{1}{(2N)^2}\sum_{p_f,q_f \in {\cal IH}_N } 
\omega_N^{2(p_fq_f-p_fq+q_fp)} \nonumber \\
&\times& \hat{P}^{-2q_f}\hat{Q}^{2p_f},  
\label{LeonFano}
\end{eqnarray}
and the marginal conditions (\ref{projcoord1}) and (\ref{projmonent1}) for this Fano operator 
become 
\begin{eqnarray}
&&\sum_{p\in {\cal IH}_N} \hat{\Delta}^{({\rm L})}(q,p)\nonumber \\
&&\;\;\;\;\;\;\;\;=\left\{
      \begin{array}{ll}
         \displaystyle{\frac{1}{N}\sum_{n \in {\cal I}_N}(\omega_N^{-q} Q)^{n}} & (q \in {\cal I}_N) \\
             0                 & (q \in {\cal H}_N)
      \end{array}\right., \;\;
\label{projcoord2}\\
&&\sum_{q\in {\cal IH}_N} \hat{\Delta}^{({\rm L})}(q,p) \nonumber \\
&&\;\;\;\;\;\;\;=\left\{
      \begin{array}{ll}
         \displaystyle{\frac{1}{N}\sum_{n \in {\cal I}_N}(\omega_N^{-p} P)^{n}} & (p \in {\cal I}_N) \\
             0                 & (p \in {\cal H}_N)
      \end{array}\right.. \label{projmonent2}
\end{eqnarray}

Instead of the transformation rule (\ref{contran}), we impose the transformation rule on operators $\hat{Q}$ and $\hat{P}$
\begin{equation}
\left\{\begin{array}{l}
\displaystyle{\hat{Q} \to \hat{U}_{\bf h}\hat{Q}\hat{U}_{\bf h}^{\dagger}=a_Q({\bf h})\hat{P}^{\lambda}\hat{Q}^{\kappa}} \\
\\
\displaystyle{\hat{P} \to \hat{U}_{\bf h}\hat{P}\hat{U}_{\bf h}^{\dagger}=a_P({\bf h})\hat{P}^{\nu   }\hat{Q}^{\mu    }}
\end{array}\right.,
\label{unitatrPQ}
\end{equation}
where {\bf h} is a $2\times 2$ matrix whose elements are given by $\kappa$, $\lambda$, $\mu$ and $\nu$ in the 
above equation 
\[
{\bf h}= \left(
      \begin{array}{cc}
         \kappa  & \mu \\
         \lambda & \nu 
      \end{array}\right),
\]
and should be an element of group ${\rm Sp}(2,\mathbb{Z})={\rm Sl}(2,\mathbb{Z})$, because the unitary transformed operators $\hat{U}_{\bf h}\hat{Q}\hat{U}_{\bf h}^{\dagger}$ and $\hat{U}_{\bf h}\hat{P}\hat{U}_{\bf h}^{\dagger}$ satisfy the same conditions as eq.(\ref{comQP}). From the same condition as eq.(\ref{PQN}), we can determine coefficients $a_Q({\bf h})$ and $a_P({\bf h})$ up to integers $N_P({\bf h})$ and $N_Q({\bf h})$,
\begin{eqnarray*}
a_P({\bf h})&=&\omega_N^{\frac{\nu   \mu     (N-1)}{2}+N_P({\bf h})}, \\
a_Q({\bf h})&=&\omega_N^{\frac{\kappa\lambda (N-1)}{2}+N_Q({\bf h})}.
\end{eqnarray*}

We choose these integers as the followings,
\begin{eqnarray}
\omega_N^{N_Q({\bf h})}
    &=&\omega_N^{-\kappa       n_{+}}
       \omega_N^{(\lambda-1)   n_{-}}, \label{solnpF} \\
\omega_N^{N_P({\bf h})}
    &=&\omega_N^{-(\mu-1)n_{+}}
       \omega_N^{ \nu n_{-}}. \label{solnsF}
\end{eqnarray}
Owing to this choice, we can show that the operators $\hat{Q}^n\hat{P}^m,\;\;(n,m \in {\cal I}_N)$  transformed by  unitary operators $\hat{U}_{\bf h}$ and $\hat{U}_{{\bf h}'}$ successively are equal to the operators transformed by the unitary operator $\hat{U}_{{\bf hh}'}$;
\[
\hat{U}_{{\bf h}'}\hat{U}_{\bf h}\hat{Q}^n\hat{P}^m\hat{U}_{\bf h}^{\dagger}\hat{U}_{{\bf h}'}^{\dagger}
=\hat{U}_{{\bf h}'{\bf h}}\hat{Q}^n\hat{P}^m\hat{U}_{{\bf h}'{\bf h}}^{\dagger}
\]
where  ${\bf h}$ and ${\bf h}'$ are arbitrary elements of the group ${\rm Sp}(2,\mathbb{Z})$.
Since any operators are described in series of operators $\hat{Q}^n\hat{P}^m$, the unitary operator $\hat{U}_{{\bf h}'}\hat{U}_{\bf h}$ is equivalent to 
$\hat{U}_{{\bf hh}'}$ up to phase factor, that is, $\hat{U}_{\bf h}$ can be considered as a projective unitary representation of the group
${\rm Sp}(2,\mathbb{Z})$,
\begin{equation}
\hat{U}_{{\bf h}'}\hat{U}_{\bf h}=e^{i\phi({\bf h},{\bf h}')}\hat{U}_{{\bf h}'{\bf h}}.
\label{cnsteq}
\end{equation}
We assume that the Fano operator is covariant under the similarity transformation with respect to the $\hat{U}_{\bf h}$;
\begin{equation}
\hat{U}_{\bf h}\hat{\Delta}(q,p)\hat{U}_{\bf h}^{\dagger}=\hat{\Delta}(\nu q-\lambda p,-\mu q+\kappa p).
\label{intcond}
\end{equation}@
Then, operator of both $\hat{U}_{{\bf h}'}\hat{U}_{\bf h}$ and $\hat{U}_{{\bf h}'{\bf h}}$ must transform the Fano operator to same operator; namely, the Fano operator with arguments $(\nu'\nu+\lambda'\mu)q-(\nu'\lambda+\lambda'\kappa)p$ and
$-(\mu'\nu+\kappa' \mu)q+(\mu'\lambda+\kappa' \kappa)p$ instead of $q$ and $p$, respectively.
This fact is ensured by eq.(\ref{cnsteq}) which is derived
by the choices (\ref{solnpF}) and (\ref{solnsF}).

In our previous paper\cite{horibe1}, we constructed the Fano operator whose arguments take integer only in the case where integer $N_P$ and $N_Q$ are equal to zero.
Using the same method as was done there, we can show that it is determined uniquely only if $n_{\pm}=0$ on odd dimensional vector space and that it does not
exist for any integer $n_{\pm}$ on even dimensional vector space.
  
In the next section, we construct the Fano $\hat{\Delta}(q,p)$ operators whose arguments $q$ and $p$ take integer and half-integer values and which fulfills two conditions, marginality and covariance.
\section{\label{sec:cnstrct}Construction of the Fano operators}
It is convenient to introduce the Fourier transformed Fano operator as follows,
\begin{eqnarray}
&&\hat{\Delta}_{F}(q_f,p_f)=\frac{1}{2N}\sum_{q,p \in{\cal IH}_N}
\omega_{N}^{2qp_f}\omega_N^{-2pq_f}\hat{\Delta}(q,p)  
 \nonumber \\
&=&
\frac{1}{2N}\sum_{q,p \in{\cal IH}_N}
\omega_{2N}^{2q\cdot 2p_f}\omega_{2N}^{-2p\cdot 2q_f}\hat{\Delta}(q,p).
\label{newfourier}
\end{eqnarray}
Here, $q_f$ and $p_f$ take integer and half-integer values, so this definition is equivalent to discrete Fourier transformation with period $2N$.
The marginality eq.(\ref{projcoord2}) and (\ref{projmonent2}) are cast into the simple form
\begin{eqnarray}
\hat{\Delta}_{F}(0,p_f)&=&\frac{1}{2N} \hat{Q}^{ 2p_f}, \label{fprojcoord} \\
\hat{\Delta}_{F}(q_f,0)&=&\frac{1}{2N} \hat{P}^{-2q_f}. \label{fprojmonent}
\end{eqnarray}
From the covariance eq.(\ref{intcond}), we get the restriction to the Fourier transformed Fano operator by explicit calculation,
\begin{equation}
\hat{U}_{\bf h}\hat{\Delta}_{F}(q_f,p_f)\hat{U}_{\bf h}^{\dagger}
=\hat{\Delta}_{F}(\nu q_f-\lambda p_f,-\mu q_f+\kappa p_f). 
\label{con4ff}
\end{equation}

Now, we find the Fourier transformed Fano operator at arbitrary point 
$(q_f,p_f),(q_f\;p_f \in {\cal IH}_N)$. 
Because the  $\hat{\Delta}_F(p_f,q_f)$ is periodic operator with period $N$ and  $\kappa$ and $\lambda$ of integer parameters in the transformation eq.(\ref{unitatrPQ}) should be relatively prime to each other because of the relation $\kappa\nu-\lambda\mu=1$, we make three integers $\kappa$, $\lambda$ and $\xi$ from $p_f$ and $q_f$ ,
\begin{equation}
\left\{\begin{array}{l}
\xi \kappa = 2p_f - 2N \left[\frac{2p_f}{N}\right],\\
\\
\xi \lambda= 2q_f - 2N \left[\frac{2q_f}{N}\right].
\end{array}\right. \label{relprime}
\end{equation}
where $\xi$ is the greatest common divisor of $2p_f - 2N \left[\frac{2p_f}{N}\right]$  and $2q_f - 2N \left[\frac{2q_f}{N}\right]$ which take values between $0$ and $2N-1$.
From eq.(\ref{con4ff}), we have 
\[
\hat{U}_{{\bf h}(q_f,p_f)}\hat{\Delta}_{F}\left(0,\frac{\xi}{2}\right)\hat{U}_{{\bf h}(q_f,p_f)}^{\dagger}
=\hat{\Delta}_{F}\left(\frac{\lambda\xi}{2},\frac{\kappa\xi}{2}\right). 
\]
where ${\bf h}(q_f,p_f)$ is an element of ${\rm Sp}(2,\mathbb{Z})$ and is given by
\[
{\bf h}(q_f,p_f)=
\left(\begin{array}{cc}
\kappa & -\mu  \\
-\lambda & \nu
\end{array}\right).
\]
where $\mu$ and $\nu$ are any integers which satisfy the relation $\kappa\nu-\lambda\mu=1$ for $\kappa$ and $\lambda$ defined by eq.(\ref{relprime}).
The right hand side can be estimated by the transformation rule eq.(\ref{unitatrPQ}) and marginality condition (\ref{fprojcoord})
\begin{eqnarray*}
&&\hat{U}_{{\bf h}(q_f,p_f)}
\hat{\Delta}_{F}\left(0,\frac{\xi}{2}\right)
\hat{U}_{{\bf h}(q_f,p_f)}^{\dagger}=
\frac{1}{2N}\hat{U}_{{\bf h}(q_f,p_f)}
 \hat{Q}^{\xi}
\hat{U}_{{\bf h}(q_f,p_f)}^{\dagger} \\
&=&
\frac{1}{2N}     
       \omega_N^{-\frac{\xi\kappa\lambda (N-\xi )}{2}}
       \omega_N^{-\xi\kappa       n_{+}}
       \omega_N^{-\xi(\lambda+1)   n_{-}}
            \hat{P}^{-\xi\lambda}\hat{Q}^{\xi\kappa},
\end{eqnarray*}
and we get
\begin{equation}
\hat{\Delta}_{F}\left(\frac{\lambda\xi}{2},\frac{\kappa\xi}{2}\right)
=
\frac{1}{2N}     
       \omega_N^{-\frac{\xi\kappa\lambda (N-\xi )}{2}-\xi\kappa n_{+}-\xi(\lambda+1)   n_{-}}
       \hat{P}^{-\xi\lambda}\hat{Q}^{\xi\kappa}.
\label{finway2}
\end{equation}

For the element ${\bf h}_{\pm}=\left(\begin{array}{cc} \pm 1 & 0 \\ \mu & \pm 1 \end{array}\right)$ of ${\rm Sp}(2,\mathbb{Z})$,
we have 
\[
\hat{\Delta}_{F}\left(0,\pm\frac{\xi}{2} \right)
=
\frac{1}{2N}     
       \omega_N^{-\xi (n_{-} \pm n_{+})}
             \hat{Q}^{\pm \xi},
\]
which is consistent with the marginality condition (\ref{fprojcoord}) if $n_{-} \pm n_{+}$ is equal to the integral multiple of $N$.
Therefore, we may get the Fano operator satisfying marginality and covariance if we choose the projective unitary representation of ${\rm Sp}(2,\mathbb{Z})$ with $n_{+}=n_{-}={\rm integer}\times N$ or with $n_{+}=n_{-}=\frac{N}{2}+{\rm integer}\times N$ in the case where $N$ is even number.

First, we study the case with $n_{+}=n_{-}=\frac{N}{2}+{\rm integer}\times N$.
Then, eq.(\ref{finway2}) becomes
\begin{eqnarray*}
\hat{\Delta}_{F}\left(\frac{\lambda\xi}{2},\frac{\kappa\xi}{2}\right)
&=&
\frac{1}{2N}     
       \omega_N^{\frac{\xi(\kappa-1)(\lambda-1)N}{2}-\xi N}
       \omega_N^{\frac{\xi^2\kappa\lambda}{2}}
       \hat{P}^{\xi\lambda}\hat{Q}^{\xi\kappa},\\
&=&\frac{1}{2N}     
       \omega_N^{\frac{\xi^2\kappa\lambda}{2}}
       \hat{P}^{-\xi\lambda}\hat{Q}^{\xi\kappa},
\end{eqnarray*}
We can obtain the second equality,  
taking account of the fact that both integers $\kappa$ and $\lambda$ are not even because of the 
condition $\kappa\nu-\lambda\mu=1$ and that $(\kappa-1)(\lambda-1)$ is an even number. Returning variables $\xi$, $\kappa$ and $\lambda$ to $p_f$ and $q_f$
by use of eq.(\ref{relprime}),
we have
\begin{equation}
\hat{\Delta}_{F}(q_f.p_f)=
\frac{1}{2N}     
       \omega_N^{2p_fq_f}
       \hat{P}^{-2q_f}\hat{Q}^{2p_f},
\label{finway0-2}
\end{equation}
which is equivalent to the Fano operator proposed by Leonhardt in
eq.(\ref{LeonFano}).
 
Second, we investigate the case with $n_{+}=n_{-}={\rm integer}\times N$. Then, equation (\ref{finway2}) becomes 
\begin{eqnarray}
\hat{\Delta}_{F}\left(\frac{\lambda\xi}{2},\frac{\kappa\xi}{2}\right)
&=&\frac{1}{2N}     
       \omega_N^{\frac{-\xi^2\kappa\lambda (N-1)}{2}}
       \omega_N^{ \kappa\lambda N \frac{\xi(\xi-1)}{2}}
       \hat{P}^{\xi\lambda}\hat{Q}^{\xi\kappa} \nonumber \\
&=&\frac{1}{2N}     
       \omega_N^{-\frac{\xi^2\kappa\lambda (N-1)}{2}}
       \hat{P}^{-\xi\lambda}\hat{Q}^{\xi\kappa}.
\label{finway1-1}
\end{eqnarray}
Since $\xi(\xi-1)$ is even number and the factor $\omega_N^{\kappa\lambda N \frac{\xi(\xi-1)}{2}}$ in the first equality is equal to unity, we get the second equality. 
Using eq.(\ref{relprime}), we can replace the variables $\xi\lambda$ and $\xi\kappa$ to $q_f$ and $p_f$, respectively and we 
obtain the operator $\hat{\Delta}_F(q_f,p_f)$ for arbitrary points $(q_f,p_f)$,
\begin{equation}
\hat{\Delta}_{F}\left(p_f,q_f\right)
=\frac{1}{2N}     
       \omega_N^{-2(N-1)p_fq_f}
       \hat{P}^{-2q_f}\hat{Q}^{2p_f},
\label{finway0-1}
\end{equation}
which is different from the Fano operator by Leonhardt only when variables
$p_f$ and $q_f$ are half-integers.
In this case, it is not used that dimension of the vector space under consideration is an even number, so far, and we can find the Fano operator for odd dimensional vector space
from the above equation, although what we want to find is the Fano operator on
even dimensional vector space. 
Let us transform the operator $\hat{\Delta}_F(p_f,q_f)$ to $\hat{\Delta}(q,p)$ by inverse
transformation of eq.(\ref{newfourier}).
Noting that the integers $2p_f$ and $2q_f$ take values between zero and $2N-1$
and that the same operator $\hat{P}^m\hat{Q}^n$ appears four times in the summation by
$p_f$ and $q_f$, we have
\begin{eqnarray}
\hat{\Delta}(q,p)&=&\frac{1}{(2N)^2}
\sum_{m,n=0}^{N-1}
\left(1+\omega_N^{-qN}\omega_N^{\frac{N(N-1)}{2}m}\right) \nonumber \\
&& \times 
\left(1+\omega_N^{ pN}\omega_N^{\frac{N(N-1)}{2}n}\right) \nonumber \\
&& \times
\omega_N^{-\frac{(N-1)}{2}nm}
\omega_N^{-qn}
\omega_N^{ pm}
\hat{P}^{-m}\hat{Q}^{n}. 
\label{finfano1} 
\end{eqnarray}
For odd dimensional vector space, eq.(\ref{finfano1}) becomes
\begin{eqnarray}
\hat{\Delta}(q,p) 
&=&\frac{1}{N^2}\sum_{m,n=0}^{N-1}
\omega_N^{-\frac{(N-1)}{2}nm}
\omega_N^{-qn}
\omega_N^{ pm} \nonumber \\
&\times&
\left\{\begin{array}{ll}
\hat{P}^{-m}\hat{Q}^{n} & (q,p \in {\cal I}_N) \\
0 &(q \notin {\cal I}_N,\;\; {\rm or}\;\; p \notin {\cal I}_N)
\end{array}\right..
\label{finfanof}
\end{eqnarray}
Thus, extension to the Fano operator on ${\cal IH}_N$ has no meaning and it reduces to the Fano operator proposed by Cohendet et al.\cite{cohen} and constructed in our previous paper\cite{horibe1}.  

\section{\label{sec:Summary}Summary and discussion}
In this note, we constructed the Fano operator which satisfies
marginality and covariance under unitary transformation like linear canonical transformation for one dimensional quantum system. We considered the projective unitary representations which are characterized by two integer-valued variables $n_{+}$ and $n_{-}$ as the unitary operator to cause this transformation.
It was shown that there are the Fano operator with these properties for only two cases. The one is for  $n_{+}=n_{-}=\frac{N}{2}+N\times {\rm integer}$ and the other is for $n_{+}=n_{-}=N\times {\rm integer}$.
For both cases, we could uniquely determine the Fano operator which is equivalent to the one given by Leonhardt in
eq.(\ref{LeonFano}) for the former case and which is new one for the latter case.  

Finally we consider the relation between ``weighted average" of function of $q$ and
$p$ with the Wigner function and expectation value of operators. 
It is sufficient to investigate  expectation  values of operators  $\hat{Q}^m\hat{P}^n$,
because any operators are expanded in series of those operators.
We use the same method as is explained for one-dimensional quantum system in the Section \ref{sec:Intro}. We start with expectation value of operator $\hat{Q}^{\xi}$.
From marginality eq.(\ref{projcoord2}), we have
\[
\sum_{q,p \in{\cal IH}_N}\omega_{N}^{q\xi}\hat{\Delta}(q,p)
=\hat{Q}^{\xi}.
\]
Applying the similarity transformation by the unitary operator $\hat{U}_{\bf h}$ to the above equation, we obtain 
\begin{eqnarray}
&&\sum_{q,p  \in {\cal IH}_N}\omega_N^{q\xi}\hat{\Delta}(\nu q-\lambda p,-\mu q+\kappa p)=\left(\hat{U}_{\bf h}\hat{Q}\hat{U}_{\bf h}^{\dagger}\right)^{\xi}, \nonumber \\
&&\sum_{q',p' \in {\cal IH}_N}\omega_N^{(\kappa  q' +\lambda p')\xi}\hat{\Delta}(q',p')
\nonumber\\
&=&\left(\omega_N^{\frac{\kappa\lambda (N-1)}{2}}
         \omega_N^{-\kappa       n_{+}}
         \omega_N^{(\lambda-1)   n_{-}}\hat{P}^{\lambda}\hat{Q}^{\kappa}\right)^{\xi}. 
\label{opor2}
\end{eqnarray}
Here, we change the integers $p$ and $q$ for summation to $p'$ and $q'$ defined by
\begin{eqnarray*}
&&\left(\begin{array}{c}
q' \\
p'
\end{array}\right)=\left(\begin{array}{cc}
\nu    & -\lambda\\
-\mu   &  \kappa 
\end{array}\right)
\left(\begin{array}{c}
q \\
p
\end{array}\right)
-\left(\begin{array}{c}
m_q \\
m_p
\end{array}\right)N \\
&&\;\;
\rightleftharpoons \;\; 
\left(\begin{array}{c}
q \\
p
\end{array}\right)
=
\left(\begin{array}{cc}
 \kappa & \lambda  \\
 \mu    &  \nu  
\end{array}\right)
\left\{\left(\begin{array}{c}
q' \\
p'
\end{array}\right)
+\left(\begin{array}{c}
m_q \\
m_p
\end{array}\right)N\right\}
\end{eqnarray*}
where $m_p$ and $m_q$ are integer part of ratios of $\nu q - \lambda p$ and $-\mu q + \kappa q$  to $N$, respectively.

Using that $\xi(\xi-1)$ and $(\kappa-1)(\mu-1)$ are even numbers, and replacing 
$\xi\kappa$ and $\xi\lambda$ by $a$ and $b$, we obtain 
\begin{eqnarray}
&&\sum_{q,p \in {\cal IH}_N}\omega_N^{( b p +a  q)}\hat{\Delta}(q,p) \nonumber \\
&=&\left\{\begin{array}{ll}
\omega_N^{\frac{ N-1}{2}ab}
     \hat{P}^{b}\hat{Q}^{a}  & (n_{\pm}=N\times{\rm integer}) \\
\omega_N^{-\frac{1}{2}ab} \hat{P}^{b}\hat{Q}^{a} &(n_{\pm}=\frac{ N}{2}+N\times{\rm integer})
\end{array}\right.,
\label{opordb}
\end{eqnarray}
that is, in term of the Wigner function, we have
\begin{eqnarray*}
&&\sum_{q,p \in {\cal IH}_N}
\left(\omega_N^{ q}\right)^a
\left(\omega_N^{ p}\right)^b
W(q,p)\\
&=&
\left\{\begin{array}{ll}
{\rm Tr}\left[\hat{Q}^{\frac{a(1-N)}{2}}\hat{P}^{b}\hat{Q}^{\frac{a(1-N)}{2}}\hat{\rho}\right] & (n_{\pm}=N\times{\rm integer}) \\
& \\
{\rm Tr}\left[\hat{Q}^{\frac{a}{2}}\hat{P}^{b}\hat{Q}^{\frac{a}{2}}\hat{\rho}\right] &(n_{\pm}=\frac{ N}{2}+N\times{\rm integer})
\end{array}\right.,
\end{eqnarray*}
where the Wigner function for density matrix $\hat{\rho}$ is defined by similar equation to eq.(\ref{wigfano}) for one-dimensional quantum system.
From eq.(\ref{defQP}), we can regard the variables $\omega_N^{ q}$ and
$\omega_N^{p}$ as classical variables corresponding to quantum operators 
$\hat{Q}$ and $\hat{P}$, respectively. So the above equation tells us the operator ordering of quantum quantities whose expectation can be calculated using the Wigner function.
We derived the eq.(\ref{opordb}) in the roundabout sort way in order to see the relation between the covariance and operator ordering. These equations are same ones as eqs.(\ref{finway0-2}) and (\ref{finway0-1}), as was pointed out in the paper
\cite{takami}. 

In this note, we have investigated the covariance of
Fano operators for the group \({\rm Sp}(2,\mathbb{Z})\).
It is also natural to consider the covariance for \({\rm Sp}(2,\mathbb{Z}_N)\)
where \(\mathbb{Z}_N\) is the cyclic group of order \(N\).
The origin of the latter group and the group theoretical treatment
of the covariance will be discussed elsewhere.

\bibliography{subwig12}

\end{document}